\documentstyle[prl,aps,preprint]{revtex}

\input BoxedEPS
\SetRokickiEPSFSpecial  %% for dvips by Tom Rokicki, for VMS
\HideDisplacementBoxes

\begin{document}
\draft
\tighten
\renewcommand{\baselinestretch}{1.4}

%\setcounter{page}{0}
%%%%%%%%%%%%%%%%%%%%%%%%%%%%%%%%%%%%%%%%%%kbb

\newcommand{\xq}{(x,Q^2)}

\title{The Double Scattering Contribution to \protect\boldmath$b_1(x,Q^2)$ in the Deuteron}

\author{K.~Bora$^{\rm a}$\footnote{On leave from Gauhati
University, India} and R.~L.~Jaffe$^{\rm a,b}$}

\address{{~}\\$^{\rm a}$Center for Theoretical Physics, 
Laboratory for Nuclear Science 
and Department of Physics \\
Massachusetts Institute of Technology,
Cambridge, Massachusetts 02139 \\ 
% and \\
$^{\rm b}$RIKEN BNL Research Center \\
Brookhaven National Laboratory, Upton, NY 11973\\
{~}}
\maketitle

%%%%%%%%%%%%%%%%%%%%%%%%%%%%%%%%%%%%%%%%%%%%%%%%%%%%%%%%%%%%%%%%%%%%%%%%%%%%%%
\centerline{Submitted to {\it Physical Review D}\qquad  hep-ph/9711323
                                        \qquad \rm MIT-CTP-2692 (Revised)}
% \preprint{\vbox{Submitted to Physical Review D
%                                         \null\hfill\rm MIT-CTP-2692 (Revised)}}
%%%%%%%%%%%%%%%%%%%%%%%%%%%%%%%%%%%%%%%%%%%%%%%%%%%%%%%%%%%%%%%%%%%%%%%%%%%%%%

%\thispagestyle{empty}

\begin{abstract} 

We study the tensor structure function $b_1\xq$ in deep inelastic
scattering (DIS) of an electron from a polarized deuteron target. We
model the electron-nucleon cross section at the starting point for
$Q^2$ evolution by vector-meson-dominance (VMD). Shadowing due to the
double-scattering of vector mesons, along with the presence of a
d-state admixture in ground state deuteron wave function gives rise to
a non-vanishing contribution to $b_1\xq$. We find a large
enhancement at low-$x$ in qualitative agreement with other recent estimates of
double-scattering contributions to $b_1\xq$.  If the model is valid, it should
apply within the range of present fixed target experiments.
\end{abstract}

\narrowtext
%%%%%%%%%%%%%%%%%%%%%%%%%%%%%%%%%%%%%
\section{INTRODUCTION}

Deep inelastic scattering from polarized targets continues to excite
interest among both theorists and experimentalists. When an electron
scatters off a spin-one target, such as a deuteron, new information
not present in case of a spin-half target can be obtained
\cite{hood89}. A new, leading twist tensor-polarized structure
function, $b_1\xq$, can be determined by measuring the cross section from a target aligned along the beam, and subtracting the cross
section for an unpolarized target. The function $b_1\xq$ vanishes if
the spin-one target is made up of spin-half nucleons at rest or in a
relative $s$-wave. In the parton model it measures the difference in
the quark momentum distributions of a helicity $1$ and $0$ target,
\begin{equation}
b_1 = \frac{1}{2}(2q^0_\uparrow - q^1_\uparrow - q^1_\downarrow),
\label{hel}
\end{equation}
where $q^m_\uparrow$ ($q^m_\downarrow$) is the probability to find a
quark with momentum fraction $x$ and spin up (down) along the
$z$-axis, in a hadron (nucleus) with helicity $m$, moving with
infinite momentum along $z$-axis.  $b_1\xq$ has not yet been measured
experimentally. Two recent papers have studied the effect of multiple
scattering on $b_1$ and found large contributions at small-$x$.  Our
aim is to explore these issues in the context of vector meson
dominance (VMD), where some of the uncertainties evident in
Refs.~\cite{niko97,edel97} are more explicit.  Within the range of
these uncertainties we find that multiple scattering does produce a
large contribution to $b_1\xq$ at small $x$. Our estimates are
smaller than those of Refs.~\cite{niko97,edel97} by factors of $1.5$
-- $2.5$, differences which are not unexpected given the conceptual differences
between their approaches and ours.  Nikolaev and Sh\"afer\cite{niko97}
use the pomeron structure function in proton to extract the
diffractive shadowing contribution.  Their results have been presented
for $Q^2=10\, {\rm GeV}^2$.  Edelmann {\it et al.\/}\cite{edel97}
estimate $b_1$ by expressing it in terms of
$\displaystyle{\frac{\delta F_2}{F_2}}$.  Our analysis does not
support such a simple scaling relation between $b_1$ and the shadowing
of $F_2$, although the two originate in the same double scattering
mechanism.  Furthermore, the authors of Ref.~\cite{edel97} do not
specify the scale at which their results should apply.  Given these
differences of approach, we view our results as qualitative
confirmation of the work of Refs.~\cite{niko97,edel97} in a specific,
rather well defined, model.

Deep inelastic scattering from nuclear targets is usually discussed in the
context of the ``convolution model'',\cite{jaffe88} where it is assumed that
the
constituents of the nucleus scatter incoherently.  An essential assumption
of the convolution model is that a quark residing inside the nucleon
absorbs the virtual photon (in a typical DIS process) while the
fragments of nucleus and the constituents propagate into the final
state without interaction or interference.  In the convolution model
$b_1$ vanishes if the $d$-state admixture in the deuteron is ignored
\cite{hood89}. The contribution to $b_1$ from the deuteron $d$-state
was studied in Ref. \cite{khan91}, along with the contribution from
double scattering from the two nucleons, which amounts to a coherent
contribution to the amplitude. In Ref.~\cite{khan91} the double scattering
process was studied at the parton level.

In the present work, we investigate the behavior of $b_1\xq$ in a
vector meson dominance (VMD) \cite{baur78} model.  Of course deep inelastic
scattering at large-$Q^2$ should be discussed in terms of quarks and
gluons. If taken literally at large $Q^2$, VMD has the wrong $Q^2$
dependence.  VMD can be used, however, to provide ``boundary value data'' ---
{\it i.e.\/} starting values for parton distribution functions --- at low-$Q^2$
where the assumptions of VMD are well-founded.  We choose VMD because it lends
itself to the treatment of multiple scattering effects that violate the
convolution model and may give rise to a significant contribution to $b_1$. 
Also, because vector meson production data are available, cross sections
necessary for our analysis can be found in the literature.  The cost of this
increased certainty is the need to identify a scale $Q_0^2$ to be assigned to
the output of the model.

Double scattering (which we will refer to as ``shadowing'') and the
$d$-state admixture in the deuteron play a crucial role in our VMD
treatment as they do in Refs.~\cite{khan91,niko97,edel97}.  According
to VMD, the virtual photon can fluctuate between a bare photon state and a
superposition of hadronic states with same quantum numbers as the
photon ($J^{PC} = 1^{--}$).  In the simplest form of VMD this state is
taken to be a superposition of $\rho$, $\omega$ and $\phi$ mesons,
\begin{equation}
\sqrt{\alpha}\left|h\right> = \sum_V\frac{e}{f_V}\frac{m_V^2}{m_V^2+Q^2}
\left|V\right>,
\label{vmd}
\end{equation}
where $em_V^2/f_V$ is the photon vector meson coupling, $\sqrt{\alpha}
\left|h\right>$ is the hadronic component of the photon, and $Q^2$ is
the virtuality of the spacelike virtual photon. As usual in VMD, we 
assume that
the vector meson interacts diffractively with the nucleon and that the
$t$-dependence of the VMD amplitudes can be taken from vector meson
photoproduction.

The VMD contribution to $b_1$ is constrained at both large and small $x$ by
simple physical effects.  For multiple scattering effects to be significant, the
time (or distance) over which the virtual vector meson can propagate through the
target nucleus (known as the ``coherence length''$\equiv\lambda$) must be long
enough for the meson to undergo more than one interaction with the target. We
shall see that the coherence length is determined kinematically by the
uncertainty principle.  At large Bjorken-$x$ ($x\ge 0.3$) $\lambda$ is smaller
than a single nucleon, so double scattering contributions to $b_1$ are
suppressed.  At small $x$, double scattering can be important.  In order to
contribute to $b_1$ it must distinguish between the helicity $\pm 1$ and
helicity $0$ states of the deuteron.  If the amplitude for $\gamma^\ast
p\rightarrow V X$ fell quickly with (transverse) momentum transfer,
corresponding to long range in impact parameter space, then shadowing could not
distinguish the orientation of the nucleons in the deuteron and $b_1$ would be
small at small $x$.  At the opposite extreme, if $\gamma^\ast p\rightarrow V X$
were flat in momentum transfer, corresponding to a $\delta$ function in impact
parameter, then shadowing would occur only when one nucleon was directly ``in
front of'' the other.  The quadrupole admixture in the deuteron wavefunction
produces just such a deformation of the wavefunction in one helicity state
relative to the other.  The deuteron is a relatively large bound state, and the
range of vector meson electroproduction is limited, so the actual situation most
closely resembles the  second scenario and leads to a significant enhancement in
$b_1$ at small $x$.\footnote{We thank G.~Piller and N.~Nikolaev for valuable suggestions on this topic.}

The paper has been organized as follows. In Section II, we present the
theoretical formulation of the model, reviewing VMD and the double
scattering analysis.  Section III contains calculations and results. 
Throughout the paper, we have attempted to keep the analysis simple and
self-explanatory.

\section {Formulation of the Model}

$b_1$ measures a tensor (spin-two) correlation of the momentum
distribution of quarks with the spin of the target in DIS.  Such a
correlation must vanish in a spin-$\frac{1}{2}$ target on account of
the Wigner-Eckart Theorem. In principle, any spin-one target
can have a non-vanishing $b_1$.  Two nucleons bound in an $s$-wave
cannot give $b_1\neq 0$.  What is not so obvious, perhaps, is that a
$d$-state admixture in a $J=1$ bound state of two nucleons will
generically give $b_1\ne 0$ 
[1,5]. It is well known that the ground
state of deutron is not spherically symmetric: it is (primarily)
an admixture of
the states $^3S_1$ ($\ell=0$,$S=1$) and $^3D_1$ ($\ell=2$,$S=1$).
This admixture produces (and was first detected through) the
observation of an electromagnetic quadrupole moment of the deuteron.
The observation of a non-zero $b_1$ through the asymmetry $b_1/F_1$
probes the same aspects of the nucleon-nucleon force as does the
deuteron's quadrupole moment.

To expose the physical significance of various structure functions, it 
is useful to describe Compton scattering in terms of helicity
amplitudes. The lepton scattering cross section from a hadron target involves
the hadron tensor 
\begin{equation}
W_{\mu\nu}(p,q,H_1,H_2)={1\over4\pi}\int d^4x \, e^{iq\cdot x}
\left<p,H_2|[J_\mu(x),J_\nu(0)]|p,H_1\right>
\label{hadtens}
\end{equation}
which is the imaginary part of the forward current-hadron scattering amplitude.
Here $H_1$ and $H_2$ are components of the target spin along a 
quantization axis and $J_\mu$ is the electromagnetic current. $W_{\mu\nu}$
can be decomposed into a set of four linearly independent structure functions
for a spin-half target using parity and time reversal invariance, while for
a spin-one target, the number of linearly independent structure function is
eight \cite{hood89}. Thus, $b_1\xq$ can be related to the helicity  
amplitudes $A_{h_1H_1,h_2H_2}$ for the process $h_1
+H_1 \rightarrow h_2 +H_2$, where  $h_j$ ($H_j$) labels the helicity
of the photon (target), and 
\begin{equation}
A_{h_1H_1,h_2H_2} = \epsilon^{\ast\mu}_{h_2}\epsilon^\nu_{h_1}W_{\mu\nu},
\label{amp}
\end{equation}
$\epsilon^\mu_h$ is the polarization vector of photon of helicity $h$. It can 
be shown that
\begin{equation}
b_1\xq = {1\over2}(2A_{+0,+0} -A_{++,++} - A_{+-,+-}).
\label{pol}
\end{equation}

The structure functions $W_1$ and $W_2$ in unpolarized DIS of an
electron from proton target can be described in terms of the
photoabsorption cross section $\sigma_T$ and $\sigma_L$ for transverse
(helicity $\pm1$) and longitudinal (helicity $0$) photons respectively
as
\begin{eqnarray}
W_1  & = & \frac{K}{4\pi^2\alpha}\sigma_T,
\label{W1}\\
W_2 & = & \frac{K}{4\pi^2\alpha}(\sigma_T+\sigma_L)\frac{Q^2}{Q^2+\nu^2}.
\label{W2}
\end{eqnarray}
$K$ is the incident virtual photon flux, $\nu$ is its energy in laboratory 
frame. 

Taking suitable combinations of helicities we can separate out
$b_1$.  We separate the contributions to $b_1$ into single and double
scattering terms.  The single scattering terms are given by the
convolution formalism and reflect the $d$-state admixture in the
deuteron ground state.\cite{hood89,khan91}.  We put these aside and
focus on the double scattering contributions, which are given by,

\begin{equation}
b_1^{(2)}\xq\ = \frac{Q^2}{8\pi^2x\alpha}\left\{\left.\delta\sigma^{(2)}_{\gamma^\ast D}
\right|_{m=0} - \left.\delta\sigma^{(2)}_{\gamma^\ast D}\right|_{m=1}\right\},
\label{lab}
\end{equation}
where
$\delta\sigma^{(2)}_{\gamma^\ast D}$ signifies the double scattering 
shadowing correction contribution to the deuteron cross section,
\begin{equation}
\sigma_{\gamma^\ast D} = \sigma^{(1)}_{\gamma^\ast p} + \sigma^{(1)}_{\gamma
^\ast n} + \delta\sigma^{(2)}_{\gamma^\ast D}.
\label{cross}
\end{equation}
In all cases $\sigma$ refers to the cross section for transverse photons ---
the subscript $ T$ has been dropped for simplicity.

\begin{figure}
\vspace*{1in}
\begin{center}
%\mbox{\psboxto(4.5in;0.0in){scar2.eps}}
\BoxedEPSF{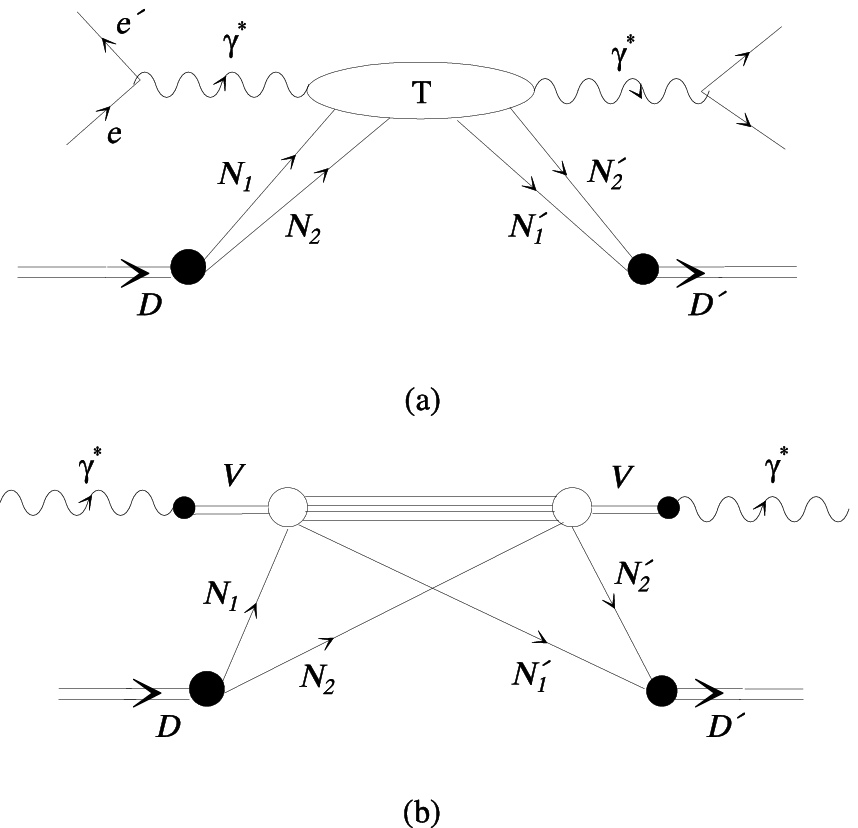 scaled 1300}
\bigskip
\caption{Double scattering diagrams in deuteron, with coherent 
contributions to amplitude from proton and neutron.} 
 \label{fig1}
\end{center}
\end{figure}

 Glauber multiple scattering theory \cite{glauber59} is usually used
to describe the interaction of high energy particles with nuclei. The
basic assumption of the Glauber treatment is that the amplitude for a
high energy hadron to interact with a nucleus can be built from the
scattering amplitude off individual nucleons. Here we shall employ
another analysis \cite{gribov72} that uses a Feynman diagram
technique, which reduces to the optical model results of Glauber
theory in the limit of large A with a general one particle nuclear
density. The double scattering contribution to photon-nucleus
scattering can be represented diagramatically as shown in
Fig.~\ref{fig1}a.  The following assumptions are made in this
analysis:~ spin and any other internal degrees of freedom are
neglected (except where necessary to isolate $b_1$), all the nucleons
are assumed to be equivalent, the momentum transfer from the incident
hadron (here the vector meson) to target nucleus is small, the nucleus
and nucleons are nonrelativistic.  The vector mesons act as the
intermediate states during the double scattering. Therefore the
double scattering picture looks as in 
Fig.~\ref{fig1}b, and the singularities of the amplitude $T$ as a
function of momenta of the intermediate vector mesons $V$ are 
isolated. They correctly correspond to the
propagation of the vector meson between nucleons. 
We fix the momentum of the virtual photon to be in $z$
direction
\begin{equation}
q^\mu = (\nu,0,0,\sqrt{Q^2+\nu^2}),
\label{photon}
\end{equation}
and define $\vec x = (\vec b, z)$.  The momentum $V^\mu$ of the vector
meson can be seen to be
\begin{equation}
V^\mu = q^\mu + P^\mu  - P'^\mu = (\nu,k_x,k_y,q_z+k_z),
\label{meson}
\end{equation}
where $P^\mu$, $P'^\mu$ are the four-momenta of initial and
final nucleon states, $N$ and $N'$ respectively, and $t=-(P-P')^2 =
k^2 $ is the momentum transfer squared. Finally, the expression for
the double scattering takes the form
\begin{eqnarray}
A^{(2)}_{\gamma^\ast D} & = & \frac{-1}{(2\pi)^3 2M}\int d^2 b\, 
dz\, |\psi(\vec b,z)|^2
\sum_V \int d^3k\, T'_{\gamma^\ast N\rightarrow NV}\times\nonumber\\
 & \times &
 \frac{e^ {i(k_zz+\vec k_\perp\cdot\vec b)}}
{\nu^2-M_V^2-{\vec k_\perp}^2-(q_z+k_z)^2 +i\epsilon} T'_{NV\rightarrow 
\gamma^\ast N},
\label{form}
\end{eqnarray}
here N,V represent the nucleon and intermediate vector meson respectively,
$T'_{\gamma^\ast N\rightarrow NV}$ is the production amplitude for vector
mesons ($\rho^0$, $\omega$, $\phi$), M is nucleon's mass and $M_V$ is the mass
of vector meson.  The amplitude $T'$ depends on the momentum transfer in
$t$- channel --- the subprocess $\gamma^\ast N\rightarrow NV$ is not limited to
the {\it forward} direction. For small $t$, $t\approx -\vec k_\perp^2$, so the
$t$-dependence determines the range of shadowing in impact parameter space. 
Even if the nucleons are misaligned in $\vec b$ space by a distance of the order
of the range of $\gamma^\ast p \rightarrow V X$, the vector meson can still
undergo a second interaction with the other nucleon.  If the range of the
production amplitude in $\vec b$ were smaller than the deuteron wavefunction,
then we
could approximate $T'$ by its value at $t=0$.  Since this is not the case we
shall have to integrate over $t$. The vector meson is not on-shell --- hence the
$i\epsilon$. The energy of the meson state is given by $E_V =
\sqrt{M_V^2+q_z^2}$ (for $\nu^2>> |k|^2$), $q_z^2 = Q^2+\nu^2$. The energy
difference that defines the virtuality of the vector meson is therefore given by
$\Delta E = \sqrt{M_V^2+\nu^2+Q^2} - \nu$, which for large values of photon
energy can be written as $\Delta E = \displaystyle{\frac{Q^2+M_V^2}{2\nu}}$. 
Therefore the
vector mesons can exist for a time $\Delta t \sim \displaystyle{\frac{1}
{\Delta E}}$, and can
propagate a distance $\lambda$, called its coherence length, $\lambda = \Delta t
= 1/\Delta E$,
\begin{equation}
\lambda = \frac{2\nu}{M_V^2+Q^2} = \frac{Q^2}{Mx(M_V^2+Q^2)}.
\label{coher}
\end{equation}
For significant shadowing or double scattering to occur, the coherence
length of the intermediate vector meson should be of order the typical
internucleon separation in the nucleus, $\sim 1.7$ fm.  Thus, these
effects increase as $x$ decreases.  Multiple scattering is most
prominant at small $x$ and for low-mass vector mesons. One can thus
justify the use of only lowest mass vector mesons in the present
case. Our model resembles partonic approaches to low-$Q^2$ shadowing
(see for example \cite{brodsky90}), where the virtual photon converts
to a $q\bar q$ pair at a distance before the target proportional to
$\displaystyle{\frac{1}{x}}$ in the laboratory frame. Shadowing is then explained in
terms of $\bar q-$nucleon scattering amplitude.  The symmetric $q\bar
q$ pairs at not too large $Q^2$, with a transverse separation $\sim
\displaystyle{\frac{1}{\sqrt{Q^2}}}$ can be viewed as a meson, the strong color
interaction between quark and antiquark increasing with increasing
separation.

\begin{figure}
\begin{center}
%\mbox{\psboxto(2.5in;0.0in){scar3.eps}}
\BoxedEPSF{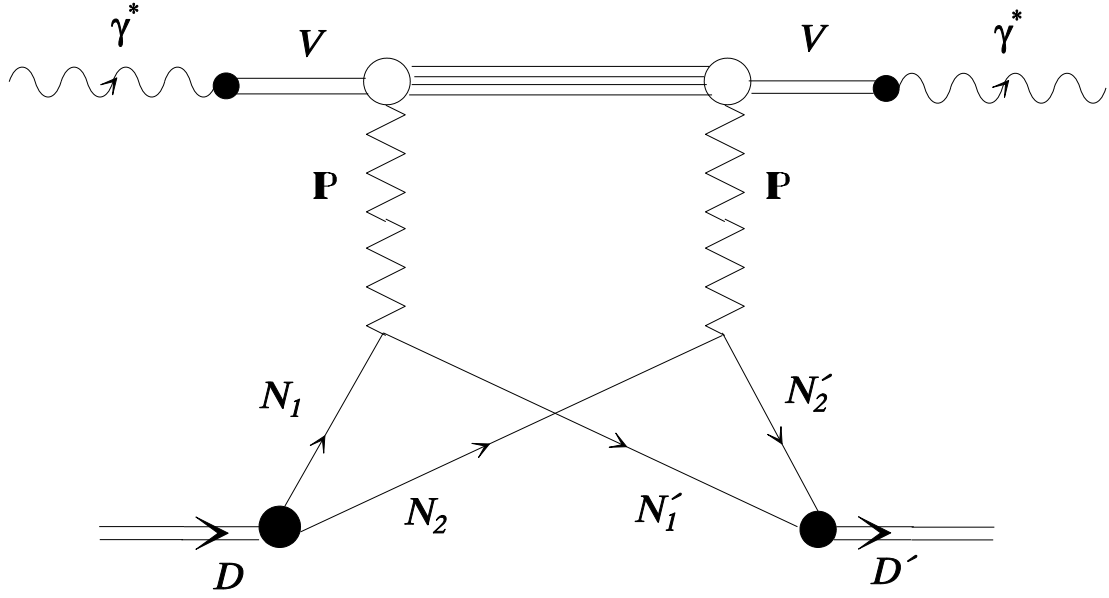 scaled 900}
\bigskip
\caption{Double scattering diagram, showing production of intermediate
vector mesons $V$ via pomeron exchange.}
 \label{fig2}
\end{center}
\end{figure}

Now we return to Eq.~(\ref{form}). The optical theorem relates the
total cross section to the imaginary part of the forward scattering
amplitude as $\delta\sigma_{\gamma^\ast D} = \frac{1}{W^2_D}
\left.{\rm Im}\, A^{(2)}_{\gamma^\ast D}\right|_{t=0}$, with $W^2_D$ the
total center of mass energy of the $\gamma^\ast$-$D$ system, $W^2_D =
2W^2_N$, $W_N^2 = (p+q)^2 \equiv 2M\nu - Q^2$.  To simplify
Eq.~(\ref{form}), we carry out the $k_z$ integration. Given the sign
of the exponential, only the singularity in upper half $k_z-$ plane
contributes.  Since the vector meson interacts diffractively with the
nucleon, the double scattering diagram looks as shown in
Fig.~\ref{fig2}. The optical theorem relates the resulting on-shell
amplitude to the differential cross section for vector meson
photoprodution,
\begin{equation}
\left.\frac{d\sigma}{dt}\right|_{t=k^2} = 
\frac{1}{16\pi}\sum_V\frac{|T'_{\gamma^\ast N
\rightarrow NV}|_{t=k^2}^2}{W_N^4},
\label{photo}
\end{equation}
where $\displaystyle{\left.\frac{d\sigma}{dt}\right|_{t=k^2\approx -{\vec
k_\perp}^2} =\left.\frac{d\sigma}{dt}\right|_{t=0}e^{-a{\vec k_\perp}^2}}$.
We estimate the $t$-dependence from photoproduction data where 
$a \approx 10.4$, $10.0$ and $7.3\, {\rm GeV}^{-2}$ for $\rho$, $\omega$
and $\phi$ vector-mesons respectively.  

Next, we consider the deuteron form factor terms in Eq.~(\ref{form}). We can 
write the deuteron wavefunction as mixture of $s$-and $d$-states ($m=1$)
\begin{equation}
\psi_{m=1} = \frac{u_0(r)}{r}Y^0_0({\Omega})\chi^1_1 + \frac{u_2(r)}{r}\left\{\sqrt{\frac{3}{5}}
Y^2_2({\Omega})\chi^{-1}_1 - \sqrt{\frac{3}{10}} Y^1_2({\Omega})\chi^0_1 + \sqrt{\frac{1}{10}}
Y^0_2({\Omega})\chi^1_1\right\},
\label{rep}
\end{equation}
where $Y$'s are the spherical harmonics and $\chi$'s are the spin wave 
functions. Using the orthogonality of the $\chi$ functions, we get

\begin{eqnarray}
|\psi_{m=1}|^2 &  =  & \frac{u_0^2}{r^2}Y^{0\ast}_0 Y^0_0 + 
\sqrt{\frac{1}{10}}
\frac{u_0u_2}{r^2}Y^{0\ast}_0 Y^0_2 + \frac{u_2^2}{r^2}\frac{3}{5}
Y^{2\ast}_2 Y^2_2 + \frac{3}{10}\frac{u_2^2}{r^2}Y^{1\ast}_2 Y^1_2\nonumber\\
&  + &
\sqrt{\frac{1}{10}}\frac{u_0u_2}{r^2} Y^{0\ast}_2 Y^0_0 + 
\frac{1}{10}\frac{u_2^2}{r^2}Y^{0\ast}_2 Y^0_2,
\label{ortho}
\end{eqnarray}
with the wave functions normalized by
\begin{equation}
\int_0^\infty dr\, [u_0^2(r) + u_2^2(r)] = 1.
\label{norml}
\end{equation}
Similarly for $m=0$,
\begin{eqnarray}
|\psi_{m=0}|^2 & =  & \frac{u_0^2}{r^2} Y^{0\ast}_0 Y^0_0 -
 \sqrt{\frac{2}{5}}
\frac{u_0u_2}{r^2} Y^{0\ast}_0 Y^0_2 + \frac{3}{10}\frac{u_2^2}{r^2}
Y^{-1\ast}_2 Y^{-1}_2 +
\frac{2}{5}\frac{u_2^2}{r^2} Y^{0\ast}_2 Y^0_2\nonumber\\ 
 & + &
\frac{3}{10}\frac{u_2^2}{r^2} Y^{1\ast}_2 Y^1_2 -\sqrt{\frac{2}{5}}
\frac{u_0u_2}{r^2} Y^{0\ast}_2 Y^0_0.
\label{and}
\end{eqnarray}
Subtracting Eq.~(\ref{ortho}) from (\ref{and}) gives
\begin{equation}
|\psi|^2_{m=0} - |\psi|^2_{m=1} = 
\frac{-3}{4\sqrt{2}\pi}\frac{u_0(r)u_2(r)}{r^2}
(3\cos^2\theta - 1).
\label{sub}
\end{equation}

Combining the results summarized in Eqs.~(\ref{lab}),(\ref{form}),
(\ref{photo}) and (\ref{sub}) the final expression for the function
$b_2^{(2)}\xq$ ($=2xb_1^{(2)}$) emerges
\begin{eqnarray}
b_2^{(2)}\xq\ & = & \frac{-3}{(\pi)^4}\frac{Q^2}{16\sqrt{2}\alpha}\,{\rm Im}\, i
\int d^2  b\,
\int dz \, u_0(r)u_2(r)
\frac{2z^2-b^2}{(z^2+b^2)^2}\times\nonumber\\
 & \times &
\sum_V \int d^2\vec k_\perp e^{iz/\lambda} e^{i\vec k_\perp\cdot\vec b
-a\vec k_\perp^2}
\frac{M_V^4}{(M_V^2+Q^2)^2} \left.\frac{d\sigma}{dt}\right|_{\gamma N
\rightarrow VN,t=0}.
\label{looks}
\end{eqnarray}
Note that the crucial quadrupole factor ($3\cos^2\theta -1$)
translates into ($2z^2-b^2$) in Eq.~(\ref{looks}).
With similar arguments, the shadowing contribution to the unpolarized
structure function can be shown to be
\begin{eqnarray}
\delta F^{(2)}_1 & = & \frac{-Q^2}{16\pi^4 \alpha}\,{\rm Im}\, i\int d^2  b\, 
\int dz \, \frac{1}{z^2+b^2}\left\{u_0^2(r) +\frac{3}{4}u_2^2(r)
\frac{b^4}{(z^2+b^2)^2}\right\}\times\nonumber\\
 & \times &
 \sum_V \int d^2\vec k_\perp e^{iz/\lambda}e^{i\vec k_\perp\cdot\vec b
-a\vec k_\perp^2}
\frac{M_V^4}{(M_V^2+Q^2)^2} \left.\frac{d\sigma}{dt}\right|_{\gamma N\rightarrow VN,t=0}.
\label{F1}
\end{eqnarray}
Since the diffractive photoproduction of vector mesons takes place via
pomeron exchange, the differential cross section for forward
scattering is of the form
$\displaystyle{\left.\frac{d\sigma}{dt}\right|_{\gamma^\ast N\rightarrow VN,
t=0}}\sim
W^{4(\alpha _P(0)-1)}$, where $\alpha_P(t=0) = 1+\delta$ is the soft
pomeron intercept.  Thus it can be seen that the scaling violations in
$b_1^{(2)}\xq$ are of the order of $\displaystyle{\frac{1}
{Q^{2(1-2\delta)}}}$, and
the contribution vanishes at large $Q^2$.  In these models, structure
function vanishes at large $Q^2$ and scaling can be restored within
the context of the model only if one takes into account the continuum
of heavier mesons (GVMD). Rather, we take the point of view that VMD
should not describe the $Q^2$ dependence because it is intrinsically a
low-$Q^2$ effective theory.  VMD provides an estimate of certain (in
this case multiple scattering) contributions to the structure function
at a low scale, which are then mapped into the large-$Q^2$ domain by
standard QCD evolution.

\section{CALCULATIONS AND RESULTS}

\begin{figure}
\begin{center}
%\mbox{\psboxto(4.5in;0.0in){bx.eps}}
\BoxedEPSF{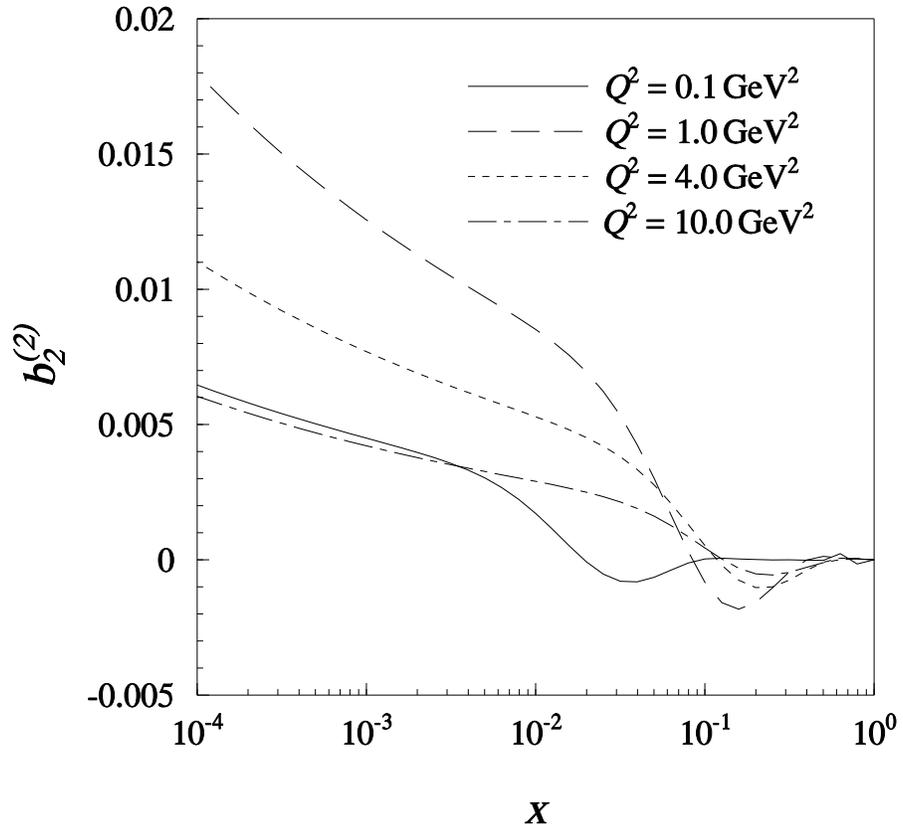 scaled 900}
\bigskip
\caption{Behavior of $b_2^{(2)}\xq$ with $x$ using Eq.(20) at $Q^2 = 0.1, 1.0, 4.0$
and $10.0{\rm GeV}^2$, with Bonn potential for deuteron.}
 \label{fig3}
\end{center}
\end{figure}

\begin{figure}
\begin{center}
%\mbox{\psboxto(4.5in;0.0in){scq.eps}}
\BoxedEPSF{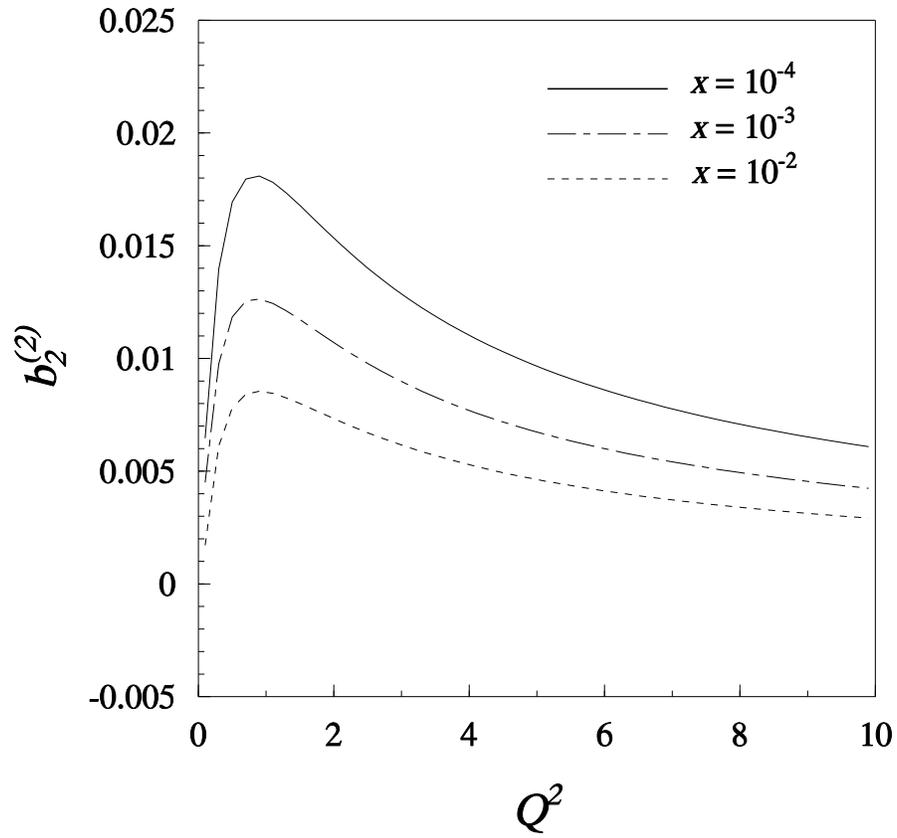 scaled 900}
\bigskip
\caption{Behavior of $b_2^{(2)}\xq$ with $Q^2$ at $x=10^{-4}$, $10^{-3}$ 
and $10^{-2}$.}
 \label{fig4}
\end{center}
\end{figure}

The resulting behavior of $b_1^{(2)}\xq$ using Eq.~(\ref{looks}) is
shown in Figs.~\ref{fig3}-\ref{fig4}. We have used the Bonn potential
\cite{mach87} for deuteron wave function in the calculations.  The
differential cross section for production of vector
mesons $\rho^0, \omega, \phi$ has been taken from reference
\cite{zeus96} and earlier data from the references therein, and have
values for forward scattering $\sim 139.0$, $10.4$ and $7.2 
\, \displaystyle{\frac{\mu b}{{\rm GeV}^2}}$ for $\rho$($W=70\, $GeV),
$\omega$($W=80\, $GeV) and
$\phi$($W=70\, $Gev) respectively. Here $W$ corresponds to the mean 
photon-proton center of mass energy. In
Fig.~\ref{fig3} we have presented the variation of $b_2^{(2)} =
2xb_1^{(2)}$ with $x$, for $10^{-4} \leq x \leq 1.0$ at $Q^2 = 0.1$, $1.0$,
$4.0$, and $10.0$ GeV$^2$. We observe that $b_2^{(2)}$ is significant
toward small $x$ values, behaving as $\sim\displaystyle{\frac{(
1-x)^{2\delta}}{x^{2\delta}}}$,
and is in general agreement with \cite{niko97,edel97}. 

\begin{figure}
\begin{center}
%\mbox{\psboxto(4.5in;0.0in){fx.eps}}
\BoxedEPSF{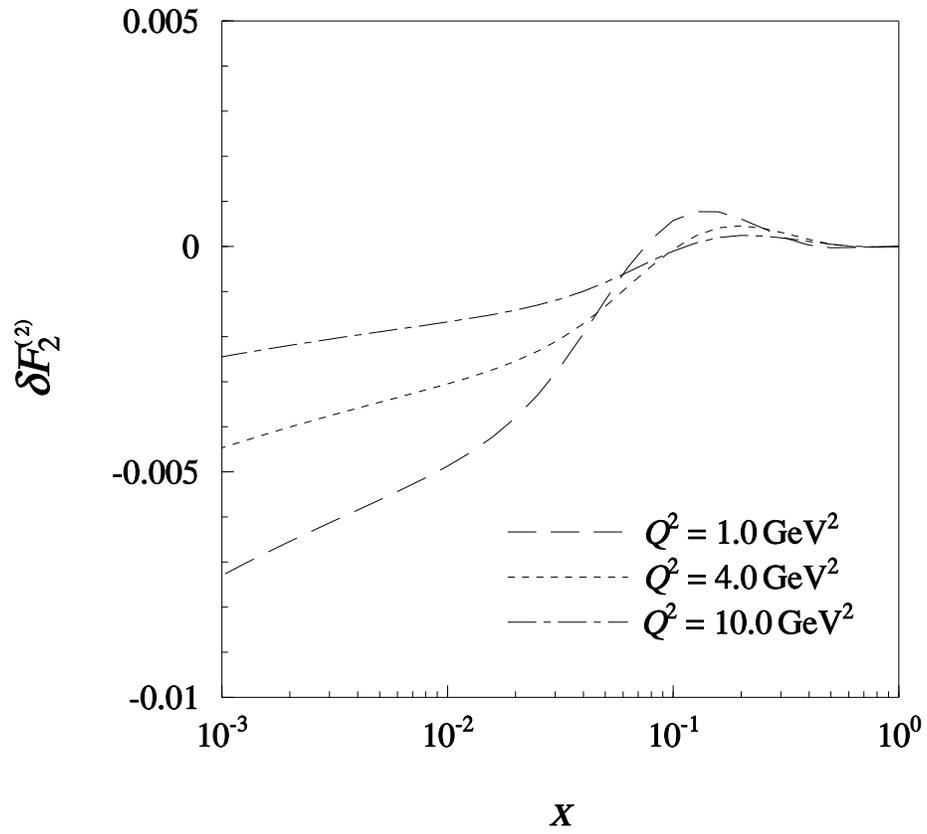 scaled 900}
\bigskip
\caption{Double scattering contribution to $F_2$ as given by Eq.(21).}
 \label{fig5}
\end{center}
\end{figure}

In Fig.~\ref{fig4} we have given the $Q^2$ behavior of $b_2^{(2)}\xq$,
as predicted by the VMD model, at different values of $x$. That
$b_1$ vanishes at $Q^2=0$ is clear from Eq.~(\ref{looks}).  It
vanishes at large $Q^2$ because the vector meson propagators and
the vector meson electroproduction cross section both fall with $Q^2$.
This can be explained by the
reduction in the coherence length of the vector mesons as $Q^2$
increases, at a fixed photon energy.  Fig.~\ref{fig5} shows the
double scattering contribution to $F_2$ in deuteron using
Eq.~(\ref{F1}).

A few comments are in order here. Our results are more specific than those of
Refs.~\cite{niko97,edel97} because we have made more specific assumptions about
the nature of the intermediate hadronic state.  Of course we could add further
excited vector mesons to our calculation, however their contribution would be
suppressed at the low $Q^2$ at which we work.  We must still confront the
question:  At which scale should we graft our VMD results onto standard QCD
evolution?  Since we are interested in qualitative rather than quantitative
behavior, some uncertainty can be tolerated.  $Q^2=0.1\, {\rm GeV}^2$ is clearly
too small --- QCD evolution is not justified at such small $Q^2$.  $Q^2=
10\,  {\rm GeV}^2$ is clearly too large --- simple vector dominance 
is not justified at
such large $Q^2$.  A choice in the range of the $\rho$ mass seems appropriate
where both VMD and QCD have claims to applicability.

To summarize, we have presented a model for the double scattering
contribution to the tensor structure function $b_1\xq$ of the
deuteron.  The analysis is based on double scattering of vector mesons
in electron-deuteron scattering. We have found that
 the double
scattering contribution to $b_1\xq$ is significant for $x\le 0.1$
and behaves as $\sim\displaystyle{\frac{(1-x)^{2\delta}}{x^{1+2\delta}}}$.
At large Bjorken-$x$ ($x\ge 0.3$) the
vector mesons can propagate only over distance scales of order the
size of a single nucleon, and multiple scattering contributions are
not significant.  At very small $x$ ($x\le 10^{-2}$), the coherence
length of the meson increases and hence the contribution increases.  Our
results agree qualitatively with those obtained in Refs.~\cite{niko97,edel97},
and confirm the fact that a significant enhancement in $b_1$ can be expected at
small $x$ due to the quadrupole deformation of the deuteron.

%%%%%%%%%%%%%%%%%%%%%%%%%%%%%%%%%%%%%%%%%%%%%%%%%%%%%%%%

%%%%%%%%%%%%%%%%%%%%%%%%%%%%%%%%%%%%%%%%%

%
%\begin{figure}
%\begin{center}
%\mbox{\psboxto(2.5in;0.0in){scar1.eps}}
%\vskip 1.0in
%\caption{The convolution model, $A$ is the nucleus, $N$ is the 
%nucleon and $q$ is constituent of nucleon.}
% \label{fig1}
%\end{center}
%\end{figure}  
%

%\newpage
%

\newpage

\newpage

\newpage

\newpage

\end{document}